\documentclass[preprintnumbers,amsmath,amssymb,
nobalancelastpage,nofootinbib]{revtex4}
\usepackage{graphicx}
\usepackage{dcolumn}
\usepackage{bm}
\usepackage{amsmath}
\usepackage{amssymb} 
\usepackage{bbm} 
\usepackage[small]{caption} 
\usepackage[small,loose]{subfigure}  
\usepackage{cancel}
\usepackage{paralist}

\newcommand{\SO}[1]{\ensuremath{\mathrm{SO}(#1)}}
\newcommand{\SU}[1]{\ensuremath{\mathrm{SU}(#1)}}
\newcommand{\U}[1]{\ensuremath{\mathrm{U}(#1)}}
\newcommand{\Z}[1]{\ensuremath{\mathbbm{Z}_{#1}}} 

\newcommand{\I}{\mathrm{i}}

\DeclareMathOperator{\Tr}{Tr}

\begin{document}

\hbox{UMD-PP-07-006; TUM-HEP-672/07}
\title{\Large  Gauged Discrete Symmetries and Proton Stability}

\author{\textbf{Rabindra N.~Mohapatra$^{a,b,c}$} and \textbf{Michael Ratz$^b$}}

\affiliation{$^a$Maryland Center for Fundamental Physics and Department of
Physics, University of Maryland, College Park, MD 20742, USA\\
$^b$Physik Department, Technische Universit\"at M\"unchen,
Garching, Germany\\
$^c$Sektion Physik, Universit\"at M\"unchen, M\"unchen, Germany
}

\date{July, 2007}

\begin{abstract}
We discuss the  results of a search for anomaly free Abelian $\Z{N}$  discrete
symmetries that lead to automatic R-parity conservation and  prevents dangerous
higher-dimensional proton decay operators in simple extensions of the minimal
supersymmetric extension of the standard model (MSSM) based on the left-right
symmetric group, the Pati-Salam group and SO(10). We require that the
superpotential for the  models have enough structures to be able to give correct
symmetry breaking  to MSSM and potentially realistic fermion masses.  We find
viable models in each of the extensions and for all the  cases, anomaly freedom
of the discrete symmetry restricts the number of generations.
\end{abstract}

\maketitle

\section{Introduction}

Supersymmetry (SUSY) is widely believed to be one of the key ingredients of
physics beyond the standard model (SM) for various reasons:
\begin{inparaenum}[(i)]
\item stability of the Higgs mass and hence the weak scale;
\item possibility of a supersymmetric dark matter;
\item gauge coupling unification, suggesting that there is
a grand unified theory (GUT) governing the nature of all forces and matter.
\end{inparaenum}
The fact that the seesaw mechanism for understanding small neutrino masses also
requires a new scale close to the GUT scale adds another powerful reason to
believe in this general picture.

There are however many downsides to SUSY. For instance, while the SM
guarantees nucleon stability, in its supersymmetric version, there  appear two
new kinds of problems: (i)~There are renormalizable R-parity breaking operators
allowed by supersymmetry and standard model gauge invariance, e.g.
\begin{equation}
W_{\cancel{R}}
~=~
\lambda_{ijk}\, L_i\,L_j\,e^c_k
+\lambda'_{ijk}\, Q_i\,L_j\,d^c_k
+ \lambda''_{ijk}\, u^c_i\,d^c_j\,d^c_k
\;.
\end{equation}
Here, $Q$, $L$, $u^c$, $d^c$ and $e^c$ denote the left-chiral quark-doublet,
lepton doublet, $u$-type, $d$-type and electron superfields.
Combination of the last two terms leads to rapid proton decay and present
limits on nucleon stability imply (for squark masses of TeV; cf.\ e.g.\
\cite{Hinchliffe:1992ad,Goity:1994dq,Dreiner:2005rd}):
\begin{equation}
\lambda'_{i1j}\,\lambda''_{11j}~\lesssim~10^{-24}\;.
\end{equation}
These terms also eliminate the possibility of any SUSY particle being the dark
matter of the Universe.\footnote{An exception to this statement is the  case
where the lightest SUSY particle is the gravitino and plays the role of the
dark  matter.} Usually assumptions such as either R-parity or matter parity
(cf.\ \cite{Dimopoulos:1981dw}) are invoked to forbid  these couplings and
rescue  the proton (as well as dark matter) stability.  (ii)~A second, more
vexing, problem is that even after imposing R-parity,  one may have dimension
five R-parity conserving operators such as
$\kappa_{ijk\ell}\,Q_i\,Q_j\,Q_k\,L_\ell/M_\mathrm{P}$.
Such operators also lead to rapid proton decay. In fact, present nucleon
stability limits put an upper limit on $\kappa_{1121},\kappa_{1122} \lesssim
10^{-8}$ \cite{Hinchliffe:1992ad}.  In what follows we will refer to these
operators as $Q\,Q\,Q\,L$.

To understand that these problems can be attributed to SUSY, recall that the
R-parity breaking terms are forbidden in the  case of SM by Lorentz invariance
in the  standard model and the $Q\,Q\,Q\,L$ is suppressed by two powers of
Planck mass and hence not problematic.

In this study, we focus on the problem of baryon number non-conservation
and require the model to satisfy the following  constraints:
\begin{inparaenum}[(i)]
\item  R-parity symmetry is exact so that it
prevents catastrophic proton decay;
\item  R-parity conserving dimension five proton decay operators of  type
$Q\,Q\,Q\,L$ are either forbidden or suppressed to  the desired  level and
\item the superpotential of the theory has enough structure  for
ensuring proper symmetry breaking down to the standard model and give
fermion masses.
\end{inparaenum}
We look for symmetries that allow all desired terms in the superpotential while
forbidding the unwanted terms discussed and satisfying the anomaly constraints.

The anomaly constraints can be related to triangle graphs involving the not only
discrete symmetry and the gauge symmetries but also gravity. The reason for
requiring anomaly freedom is the following: there is a general belief that
non-perturbative gravitational effects  such as black holes and worm  holes
break global symmetries of  nature~\cite{Giddings:1987cg}. In an
effective  field
theory language, they are  parametrized by Planck scale suppressed higher
dimensional operators. In fact, the above-mentioned $Q\,Q\,Q\,L$ type operators
are a manifestation of the fact that  global baryon number symmetry is broken by
such effects. On the other hand no hair theorem of general relativity says that
any non-perturbative gravitational effect must respect the gauge  symmetries.
Therefore if there is a gauged discrete symmetry in  the theory that prevents
the undesirable terms under discussion, they will be absent even after all
non-perturbative effects are taken into account.

How does one ensure that a discrete symmetry is a gauge symmetry\,?  This
problem has been extensively studied in the literature in the context of  the
MSSM~\cite{Krauss:1988zc,Ibanez:1991hv,Hinchliffe:1992ad,Banks:1991xj,Babu:2002tx,Babu:2003qh,Dreiner:2005rd}
and the general procedure is to calculate the various anomaly equations
involving the discrete group with gravity and the gauge group i.e.\ vanishing of
$D\,g\,g$, $D\,G^2$ anomalies where $D$ stands for the discrete
symmetry group in question, $g$ is the graviton and $G$ is the (continuous)
gauge symmetry on which the theory is based.

In the context of the MSSM, a discrete \Z6 symmetry has been identified, dubbed
`proton-hexality' in \cite{Dreiner:2005rd}, that contains  R-parity as \Z2
subgroup and forbids $Q\,Q\,Q\,L$ \cite{Ibanez:1991hv}. Remarkably, anomaly
freedom of this \Z6 requires the number of generations to be 3
\cite{Hinchliffe:1992ad}. Moreover, it has been shown that (given 3
generations) this is the only anomaly free symmetry that allows the  MSSM
Yukawa couplings and neutrino masses while forbidding the dangerous operators
\cite{Dreiner:2005rd}. Such symmetries can be extended such as to also forbid
the $\mu$ term \cite{Kurosawa:2001iq}. (For an approach to ensure proton
stability by flavor symmetries see e.g.~\cite{Kajiyama:2005rk}.) On the other
hand, the charge assignment is  different for different standard model
representations. This raises the question whether nucleon stability can be
ensured by discrete symmetries in (unified) theories where standard model
representations get combined in larger multiplets, and the charge assignment is
hence  restricted more strongly. We therefore seek  discrete symmetries ensuring
sufficient proton stability in three gauge extensions of the supersymmetric
standard model:
\begin{inparaenum}[(i)]
\item the left-right symmetric model~\cite{Senjanovic:1975rk},
\item the Pati-Salam model~\cite{Pati:1974yy} and
\item SO(10).
\end{inparaenum}
All these models incorporate the $B-L$ gauge group, which is generally used in
the discussion of the  seesaw mechanism for neutrino
masses~\cite{Minkowski:1977sc,Gell-Mann:1980vs,Yanagida:1980,Glashow:1979vf,Mohapatra:1980ia}
(for a review see e.g.\ \cite{Mohapatra:2006gs}) and also provides one way to
guarantee R-parity
conservation~\cite{Mohapatra:1986su,Font:1989ai,Martin:1992mq}. Due to the
higher gauge symmetry, which must be broken down to the SM  gauge symmetry,
specific terms must be present in the superpotential. This poses constraints
on the discrete symmetry.
%

One main result of the study is a connection between the order of the  discrete
group and the number of generations in all the cases.  We give examples of
viable models for all the different gauge groups. In our discussion, we follow a
certain `route of unification'. We start with the left-right  symmetric
extension of the supersymmetric standard model, proceed via the Pati-Salam model
to SO(10) GUTs and finally comment on how our results might be used in string
compactifications.

This paper is organized as follows: in section~\ref{sec:LR}, we discuss
anomaly-free gauge symmetries ensuring proton stability in left-right models; we
proceed to the Pati-Salam model in section~\ref{sec:PatiSalam}, and   in
section~\ref{sec:SO10} we discuss SO(10) models. We give our conclusions in
section~\ref{sec:summary}.


\section{Left-right model and discrete symmetries}
\label{sec:LR}

In this section, we discuss the left-right symmetric extension of MSSM, i.e.\
the gauge group is $\SU3_c\times \SU2_\mathrm{L}\times \SU2_\mathrm{R}\times
\U1_{B-L}$. This amounts to  breaking $B-L$ symmetry of the model by either a
$B-L=2$ triplet or a $B-L=1$ doublet. In  Table~\ref{tab:LR} (a), we show the
assignments of the quarks and  leptons and Higgs bosons. Tables~\ref{tab:LR} (b)
and (c) list the Higgs sectors of the doublet and triplet models, respectively.

\begin{center}
\begin{table}[h]
\subtable[MSSM part.]{\begin{tabular}{|c|c|}
\hline
Field           & quantum numbers \\
\hline
 $Q$            & $(\bf{3},\bf{2}, \bf{1},
+\frac{1}{3})$          \\
 $Q^c$          & $(\overline{\bf{3}},\bf{1},
\bf{2}, -\frac{1}{3})$          \\
 $L$            & $(\bf{1},\bf{2}, \bf{1},
-1)$                    \\
 $L^c$          & $(\bf{1},\bf{1}, \bf{2},
+1)$                    \\
 $\Phi$         & $(\bf{1},\bf{2}, \bf{2},
0)$                 \\
\hline
\end{tabular}}
\quad
\subtable[Doublet model.]{
\begin{tabular}{|c|c|}
\hline
Field           & quantum numbers \\
\hline
 $\chi^c$       & $(\bf{1},\bf{1}, \bf{2},
+1)$
\\
 $\overline{\chi}^c$    & $(\bf{1},\bf{1}, \bf{2},
-1)$                    \\
  $\chi$        & $(\bf{1},\bf{2}, \bf{1},
-1)$
\\
$\overline{\chi}$   & $(\bf{1},\bf{2},
\bf{1}, +1)$                    \\
$S$ & $(\bf{1},\bf{1}, \bf{1}, 0)$
            \\
\hline
\end{tabular}}
\quad
\subtable[Triplet model.]{
\begin{tabular}{|c|c|}
\hline
Field           & quantum numbers \\
\hline $\Delta^c$      &  $(\bf{1},\bf{1}, \bf{3}, -2)$
\\
$\overline{\Delta}^c$   &  $(\bf{1},\bf{1},
\bf{3}, +2)$                    \\
$\Delta$        &  $(\bf{1},\bf{3}, \bf{1}, +2)$
\\
$\overline{\Delta}$ &  $(\bf{1},\bf{3},
\bf{1}, -2)$                    \\
\hline
\end{tabular}}
 \caption{Assignment of the fermion and Higgs fields to the  representation of
the left-right symmetry group  $\SU3_c\times\SU2_\mathrm{L} \times
\SU2_\mathrm{R}  \times
\U1_{B-L}$. (a) shows the MSSM sector, (b) shows  the Higgs sector of the
doublet model and (c) the Higgs sector of the triplet model.}
\label{tab:LR}
\end{table}
\end{center}

\subsection{Left-right symmetric models -- doublet Higgs case}

The (wanted) superpotential is given by
\begin{equation}
\label{eq:SuperWTerms}
\begin{aligned}[b]
W~ =~&  \I\, h\, Q^T\, \tau_2\, \Phi\, Q^c
        + \I\, h^\prime\, L^T\, \tau_2\, \Phi L^c\\
&{}+ \I\, f_c\, L^{cT}\,
\tau_2\,\overline{\chi^c}\, \overline{{\chi^c}}^T\,\tau_2 L^c
 + \I\, f_c L^{T} \tau_2 \overline{\chi} L^T\tau_2\overline{\chi}\\
 &{}+ S\,(\chi^c\,\overline{\chi^c}+\chi\,\overline{\chi}-v^2_\mathrm{R})
 + \mu\, \Tr(\Phi^2)\;.
\end{aligned}
\end{equation}
On the other hand, the following couplings must be forbidden
(we suppress coefficients):
\begin{eqnarray}
W_\mathrm{unwanted} & =& Q^3\,L+{Q^c}^3\,L^c +Q^3\,\chi+{Q^c}^3\,\chi^c
                 +L\,\overline{\chi}+L^c\,\overline{\chi}^c
+   L\,Q\,Q^c\,\chi^c+Q\,\chi\, Q^c\, L^c
\nonumber\\
& & {}
+    L^2\, L^c\,\chi^c
+
\overline{\chi}\,L\,\Phi^2+\overline{\chi}^c\,L^c\,\Phi^2+L\,\Phi\,\chi^c
+L^c\,\Phi\,\chi\;.
\label{eq:Wunwanted}
\end{eqnarray}

To study the anomaly constraints in this model for an arbitrary $\Z{N}$
group, we start by giving the charge assignments under $\Z{N}$ to the
various superfields (denoted as $q_F$ for the field $F$ in the equation
below) and writing down the anomaly constraints.
The anomaly constraints for $\Z{N}\,g\,g$, $\Z{N}\,[\SU3_c]^2$,
$\Z{N}\,[\SU2_\mathrm{L}]^2$ and $\Z{N}\,[\SU2_\mathrm{R}]^2$ respectively
are\footnote{We ignore cubic constraints (cf.\ \cite{Banks:1991xj}).}
\begin{subequations}\label{eq:AnomalyConstraints}
\begin{eqnarray}
  N_g\,\left[6\,(q_Q+q_{Q^c})+2\,(q_L+q_{L^c})\right] +4\,q_\Phi
  +2\,(q_{\chi}+q_{\chi^c}+q_{\overline{\chi}}+q_{\overline{\chi}^c})
& = & 0\mod N'\;,\\
 N_g\,\left[2\,(q_Q+q_{Q^c})\right]
 & = & 0\mod N\;,\\
 N_g\,
 \left[3\,q_Q+q_{L}\right]+2\,q_\Phi+q_{\chi}+q_{\overline{\chi}}
 & = & 0\mod N\;,\\
 N_g\,
 \left[3\,q_{Q^c}+q_{L^c}\right]+2\,q_\Phi+q_{\chi^c}+q_{\overline{\chi}^c}
 & = & 0\mod N\;,
\end{eqnarray}
\end{subequations}
where $N_g$ denotes the number of generations.
In the first equation
\begin{equation}
 N'~=~\left\{\begin{array}{ll}
  N\;,\quad & N\:\text{odd}\;,\\
  N/2\;,\quad & N\:\text{even}\;,
 \end{array}\right.
\end{equation}
which follows from equation (10) of \cite{Ibanez:1991hv}.

The assignments must be consistent with the superpotential
\eqref{eq:SuperWTerms} and has to forbid the terms in \eqref{eq:Wunwanted}.
We scanned over possible non-anomalous $\Z{N\le12}$ symmetries, keeping
the number of
generations $N_g$ as a free parameter. Remarkably, the smallest viable
$N_g$ we
found is 3, and the smallest $N$ that works with 3 generations is $6$. An
example is shown in Table~\ref{tab:Z6chargesLRdoublet}.

\begin{table}[h]
\centerline{
\begin{tabular}{|c|c|c|c|c|c|c|c|c|c|}
\hline
$q_L$ & $q_Q$ & $q_{L^c}$ & $q_{Q^c}$ & $q_\Phi$ & $q_\chi$ & $q_{\chi^c}$ &
$q_{\overline{\chi}}$ & $q_{\overline{\chi}^c}$ & $q_S$\\
\hline
$1$ & $1$ & $5$ & $5$ & $0$ & $0$ & $2$ & $0$ & $4$ & $0$\\
\hline
\end{tabular}}
\caption{A viable \Z6 charge assignment in the left-right model with doublets.}
\label{tab:Z6chargesLRdoublet}
\end{table}

 By giving vacuum expectation values (vevs) to the fields $\chi^c$ and
$\overline{\chi}^c$, the \Z6 symmetry
is broken to a \Z2 symmetry under which matter is odd while the MSSM Higgs are
 even. That is, we have obtained an effective R-parity which, although
there is a
gauged $B-L$ symmetry, originates from an `external' \Z6.

\subsection{Left-right models -- triplet case}

The transformation properties of the fields under the gauge group are shown in
Table~\ref{tab:LR} (a) and (c). In this case, the right handed neutrino masses
arise from the  renormalizable couplings in the theory.  We have to forbid
$Q^3\,L$ and $(Q^c)^3\,L^c$. There are many anomaly-free discrete symmetries
which do the job. The interesting  point is that in  this case, the minimum
number of generations is $N_g=2$  with $N=2$ for the discrete  group. The
smallest symmetry that works for 3 generations is \Z3.

\section{Pati-Salam Model}
\label{sec:PatiSalam}

We now proceed to the Pati-Salam model, i.e.\ the gauge group is
$G_\mathrm{PS}=\SU4_c\times \SU2_\mathrm{L}\times \SU2_\mathrm{R}$. The new
feature of this model compared to the left right model just  discussed is that
the quarks and leptons belong to the same representation  (see Table
\ref{tab:PatiSalam}).
 \begin{center}
\begin{table}[h]
{\begin{tabular}{|c|c|}
\hline
Field                   & quantum numbers \\
\hline
 $\psi$                    & $(\bf{4},\bf{2},\bf{1})$\\
$\psi^c$    &$(\bf{\bar{4}}, \bf{1}, \bf{2})$\\
 $\Phi$                 &
 $(\bf{1},\bf{2}, \bf{2})$  \\
\hline
\end{tabular}}
\quad
{\begin{tabular}{|c|c|}
\hline
Field                   & quantum numbers \\
\hline
 $\chi^c$               & $(\bf{4},\bf{1}, \bf{2})$
\\
 $\overline{\chi}^c$    & $(\bf{\bar{4}},\bf{1},
\bf{2})$\\

 $\chi$                & $(\bf{4},\bf{2}, \bf{1})$
\\
$\overline{\chi}$       & $(\bf{\bar{4}},\bf{2},
\bf{1})$                                    \\
$S$     & $(\bf{1},\bf{1}, \bf{1})$
                        \\
\hline
\end{tabular}}
\quad
{\begin{tabular}{|c|c|}
\hline
Field                   & quantum numbers \\
\hline $\Delta^c$              &  $(\bf{10},\bf{1}, \bf{3})$
\\
$\overline{\Delta}^c$   &  $(\bf{\overline{10}},\bf{1},
\bf{3})$                                    \\
$\Delta$                &  $(\bf{10},\bf{3}, \bf{1})$
\\
$\overline{\Delta}$     &  $(\bf{\overline{10}},\bf{3},
\bf{1})$                                    \\
\hline
\end{tabular}}
 \caption{Assignment of the fermion and Higgs fields to the
representation of
the left-right symmetry group  $\SU4_c\times\SU2_\mathrm{L} \times
\SU2_\mathrm{R} $: (a) shows the MSSM sector, (b) shows the Higgs sector
of the doublet model and (c) the Higgs sector of the triplet model.}
\label{tab:PatiSalam}
\end{table}
\end{center}

The rest of the discussion is completely analogous to the left-right case  and
in fact the solutions displayed in Table II for the discrete charges apply to
the doublet case, and as in the triplet version of the left-right model there
are many solutions. Note that the triplet version of Pati-Salam model
does not have proton decay due to $\SU{3}_c\subset \SU4$ invariance.

\section{SO(10) GUT and discrete symmetries}
\label{sec:SO10}

\begin{figure}
\centerline{\subfigure[{}]{\includegraphics{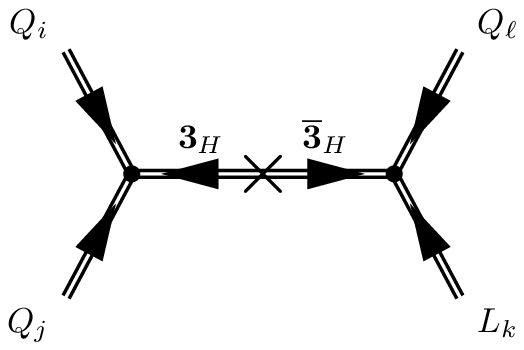}}
\qquad
\subfigure[{}]{\includegraphics{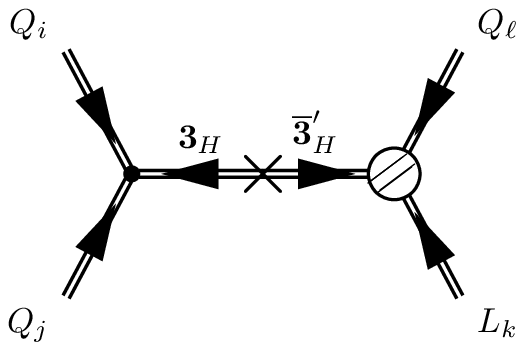}}}
\caption{Effective proton decay operators. (a)~shows the usual triplet exchange
diagram~\cite{Sakai:1981pk,Weinberg:1981wj}. (b)~illustrates that the amplitude vanishes if the mass partner of the
Higgs triplet does not couple to SM particles~\cite{Babu:1993we}.}
\label{fig:ProtonDecay}
\end{figure}

We now turn our attention to the discussion of discrete symmetries in
SO(10) GUTs.
In SO(10) models, the dimension 5 proton decay operators
$Q\,Q\,Q\,L$ have two sources:\begin{inparaenum}[(i)] \item the
higher-dimensional coupling $[\bf{16}_m]^4$ and \item effective
operators emerging from integrating out Higgs
triplets~\cite{Sakai:1981pk,Weinberg:1981wj} (see
figure~\ref{fig:ProtonDecay}). Triplet masses of the order
$M_\mathrm{GUT}$ appear to be too small to be consistent with the
observed proton life time \cite{Dermisek:2000hr}.
\end{inparaenum}
As discussed, the coefficient of the $Q\,Q\,Q\,L$ operators have to be strongly
suppressed.  This requires an explanation of why both contributions to these
operators are simultaneously small. As we shall see, such an explanation might
arise from a simple discrete symmetry.

The proton decay via Higgs triplet exchange can be forbidden by
eliminating the mass term for the \textbf{10} multiplet ($H$).
However one  needs to  make the color triplets in $H$ heavy so
that coupling unification is  maintained. This can be done by
introducing a second $\bf{10}$-plet ($H'$) such that it forms a
mass term with   $H$ but the color triplet field in it  does not
couple to standard model matter \cite{Babu:1993we}. Therefore,
from now on we will consider \SO{10} models with 2
\textbf{10}-plets.

We will discuss two classes of models:
\begin{inparaenum}[(i)]
\item where $B-L$ is broken  by \textbf{16}-Higgs fields  (cf.\
Table~\ref{tab:SO10} (b)) and
\item where $B-L$ is broken by \textbf{126}-Higgs fields~\cite{Clark:1982ai,Aulakh:1982sw}
(cf.\ Table~\ref{tab:SO10} (c)).
\end{inparaenum}
We consider Abelian discrete ($\Z{N}$) symmetries  and require
that higher dimensional R-parity conserving leading  order $\Delta
B\neq 0$ operators, which in this case are of type $\bf{16}^4_m$
(where the subscript $m$ stands for matter), are forbidden as are
all R-parity breaking terms while allowing all terms in the
superpotential that are needed to break the GUT symmetry down to
the MSSM. Our focus is on proton stability, and we leave other
issues such as fermion masses and doublet-triplet splitting for
future studies.

\begin{table}[h]
\centerline{
\subtable[MSSM part.]{\begin{tabular}{|c|c|}
\hline
Field & quantum numbers\\
\hline
$\psi_m$ & $\bf{16}_1$\\
$H$ & $\bf{10}_{-2}$\\
$H'$ & $\bf{10}_{2}$\\
\hline
\end{tabular}}
\quad \subtable[$\bf{16}$-Higgs model.]{\begin{tabular}{|c|c|}
\hline
Field & quantum numbers\\
\hline
$\psi_H$ & $\bf{16}_{-2}$\\
$\overline{\psi}_H$ & $\overline{\bf{16}}_{2}$\\
$A$ & $\bf{45}_{0}$\\
$S$ & $\bf{54}_{0}$\\
\hline
\end{tabular}}
\quad \subtable[$\bf{126}$-Higgs model.\!]{\begin{tabular}{|c|c|}
\hline
Field & quantum numbers\\
\hline
$\Delta$ & $\bf{126}_{2}$ \\
$\overline{\Delta}$ & $\overline{\bf{126}}_{-2}$ \\
$\Phi$ & $\bf{210}_{0}$ \\
\hline
\end{tabular}}}
\caption{SO(10) model. (a) shows the MSSM sector, (b) the Higgs
content of the $\bf{16}$-Higgs model and (c) the Higgs content of
the $\bf{126}$-Higgs model. \Z6 charges (see text) appear as
subscript.} \label{tab:SO10}
\end{table}

\subsection{\textbf{16}-Higgs models}

In this class of models, one has an independent motivation for
introducing a second \textbf{10}-plet coming from doublet-triplet
splitting. Further, apart from a pair of
$\bf{16}\oplus\overline{\bf{16}}$-Higgses, \textbf{45}- and
\textbf{54}-plets are required to ensure proper GUT symmetry
breaking down to MSSM
(cf.~\cite{Babu:1998wi,Albright:1998vf,Albright:1997xw,Ji:2005zk}).
The superpotential terms that must be allowed are: $\psi_m\,\psi_m\, H$,
$(\psi_m\,\overline{\psi}_H)^2/M_\mathrm{P}$, $\psi_H\,\overline{\psi}_H$,
$A^2$,  $S^{2,3}$ and $S\,A^2$, where $\psi_{m,H}$ are matter and Higgs
\textbf{16}-plets, $H$, $A$, $S$ are \textbf{10}-, \textbf{45}-,
\textbf{54}-plets, respectively (see tables~\ref{tab:SO10} (a) and (b)).
This leads to the following constraints on the \Z{N} charges:
\begin{subequations}\label{C1}
\begin{align}
2q_{\psi_m}+q_{H} & = ~ 0\mod N\;, & \quad
2q_{\psi_m}+2q_{\overline{\psi}_H} & = ~ 0\mod N\;, \\
q_{\psi_H}+q_{\overline{\psi}_H} & = ~ 0\mod N\;,& \quad
q_{H}+q_{H'} & = ~ 0\mod N\;,\\
2q_A & = ~ 0\mod N\;,& \quad
2q_S & = ~ 0\mod N\;,\\
3q_S & = ~ 0\mod N\;,& \quad
2q_A+q_S & = ~ 0\mod N\;.
\end{align}
\end{subequations}
Here, we denote the $\Z{N}$ charge for a field $F$ by $q_F$.
Next, we list the anomaly constraints,
\begin{subequations}\label{C2}
\begin{eqnarray}
\quad 16\,(N_g\, q_{\psi_m}+q_{\psi_H}+q_{\overline{\psi}_H})
+10\,(q_{H}+q_{H'})+45\,q_{A}+54\,q_{S}&=&0\mod N'\;\\
2N_g\,q_{\psi_m}+2\,q_{\psi_H}+2\,q_{\overline{\psi}_H}+q_{H}+q_{H'}+
8q_{A}+12q_{S}&=&0\mod N\;.
\end{eqnarray}
\end{subequations}
To forbid the dangerous couplings $\psi_m\,\psi_m\,H'$, $\psi_m^4$,
$\psi_m\,\overline{\psi}_H$ and $\psi_m^3\,\psi_H$,  $\psi_m\,\psi_H\,H$,
$\psi_m\,\psi_H\,H'$ and $\psi_m\overline{\psi}_HA$,
the values of the $\Z{N}$
charges have to be chosen such that they satisfy the inequalities
\begin{subequations}\label{C3}
\begin{align}
 2\,q_{\psi_m}+q_{H'} & \neq ~ 0\mod N\;,& \quad
 4\,q_{\psi_m}& \neq ~ 0\mod N\;, \\
 q_{\psi_m}+q_{\overline{\psi}_H} & \neq ~ 0\mod N\;,& \quad
 3\,q_{\psi_m}+q_{\psi_H}& \neq ~ 0\mod N\;,\\
q_{\psi_m}+q_{H',H}+q_{\psi_H} & \neq ~ 0\mod N\;,& \quad
q_{\psi_m}+q_{\overline{\psi}_H}+q_A & \neq ~ 0\mod N\;.
\end{align}
\end{subequations}
The smallest symmetry allowing to fulfill all criteria is \Z6 and requires
$N_g=3$. A possible charge  assignments is $q_{\psi_m}=1$, $q_{\psi_H}=-2$,
$q_{\overline{\psi}_H}=+2$, $q_{H}=-2$, $q_{H'}=+2$, $q_{45}=0$ (cf.\
tables~\ref{tab:SO10} (a) and (c)). This charge  assignment allows for seesaw
couplings and the possibility of fermion masses from couplings of type
$\psi_m\,\psi_m\, H$. The allowed operator $\psi_m\,\psi_m\,
\overline{\psi}_H^2$ contributes to both  the fermion masses as well as to the
seesaw. We note that the charge assignment is such that we have
\begin{equation}
 3\times\text{generation}+\text{vector-like}
\end{equation}
under $\SO{10}\times\Z6$.
The model also eliminates the dangerous proton decay operator
$Q\,Q\,Q\,L$ or operator of type $(\psi_m)^4/M_\mathrm{P}$.
We note that, although there are higher-dimensional gauge (and $\Z6$)
invariant operators, they do not give rise to
proton decay operators for the following reasons:
\renewcommand{\labelenumi}{(\roman{enumi})}
\begin{enumerate}
\item any possible combination of $\psi_H$ and $\overline{\psi}_H$ that breaks
$B-L$ cannot multiply the $B-L$ neutral proton decay operators $Q\,Q\,Q\,L$ (or
$u^c\,u^c\,d^c\,e^c$),
\item any $B-L$ neutral combination is also \Z6 invariant because it has to
involve as many $\psi_H$ as $\overline{\psi}_H$ fields, as only the SM singlets
with $B-L$ charge $\pm1$ attain a vev. Therefore the product of the \Z6
non-invariant combination $Q\,Q\,Q\,L$ with the $B-L$ neutral combination of
$\psi_H$ and $\overline{\psi}_H$ fields cannot be \Z6 invariant.
\end{enumerate}
Consider, for instance, the operators
$\frac{1}{M_\mathrm{P}^5}[\psi_m]^4\,[\overline{\psi}_H]^4$ and
$\frac{1}{M_\mathrm{P}^3}[\psi_m]^4\,[\psi_H\,\overline{\psi}_H]$. In the first
operator, the expectation value of $[\psi_H]^4$ vanishes because of the first
argument while the second operator is not \Z6 invariant.

We also observe that operators like $\psi_H^4\,H^2$, which would lead to small
diagonal entries in the $H-H'$ mass matrix, do not exist for the same reason.
That is, in this model, proton decay is only due to dimension 6 operators.

We refrain from spelling out the detailed phenomenological analysis of this
model. However, our preliminary studies seem to indicate that one can achieve
doublet-triplet splitting and realistic fermion masses while avoiding proton
decay problem by extending the Higgs content. A complete analysis of these
issues are defered to a future publication.

Let us also comment that, like in the left-right model with doublets, the \Z6
symmetry gets broken by the $\psi_H$ and $\overline{\psi}_H$ vevs down to a \Z2
which forbids $W_{\cancel{R}}$, i.e.\ acts as an R-parity. This means that
R-parity in this \SO{10} model does not originate from $B-L$.

\subsection{\textbf{126}-Higgs models}

We now discuss models where the $\bf{16}_H\oplus
\overline{\bf{16}}_H$ get replaced by
$\bf{126}\oplus\overline{\bf{126}}$ -- the motivation being that
R-parity becomes an automatic symmetry. Such models have been
extensively discussed in the
literature~\cite{Babu:1992ia,Fukuyama:2002ch,Bajc:2002iw,Goh:2003sy}.
In our context it means that the last two of the four inequalities
in Eq.~\eqref{C3} do not exist (see Eq.~\eqref{C6} below). Instead
we have the  following set of constraints on the charges from
anomaly freedom
\begin{subequations}\label{C4}
 \begin{eqnarray}
16\,N_g\, q_{\psi_m}+10\,(q_{H}+q_{H'})
+126\,(q_{\Delta}+q_{\overline{\Delta}})&=&0\mod N'\;, \\
2\,N_g\,q_{\psi_m}+35\,(q_{\Delta}+ q_{\overline{\Delta}})+
q_{H}+q_{H'}&=&0\mod N\;.
\end{eqnarray}
\end{subequations}
The superpotential constraints  can be decomposed in analogs of
Eq.~\eqref{C1}
\begin{subequations}\label{C5}
\begin{eqnarray}
2q_{\psi_m}+q_{H}& = & 0\mod N\;,\\
2q_{\psi_m}+q_{{\overline{\Delta}}} & = & 0\mod N\;,\\
q_{\Delta}+q_{\overline{\Delta}}& = & 0\mod N\;,
\end{eqnarray}
\end{subequations}
and analogs of Eq.~\eqref{C3}
\begin{subequations}\label{C6}
\begin{eqnarray}
 2\,q_{\psi_m}+q_{H'} & \neq & 0\mod N\;,\\
 4\,q_{\psi_m}& \neq & 0\mod N\;.
\end{eqnarray}
\end{subequations}

Typically in a class of these models, there are only \textbf{210} dimensional
representations that need to couple to $\Delta$ and $\overline{\Delta}$ fields
among themselves as well as with \textbf{10}
Higgs~\cite{Bajc:2004xe,Fukuyama:2004ti}. These imply  that the discrete charge
of \textbf{210} vanishes and also that of $\Delta$ and $\overline{\Delta}$ are
opposite. Substituting these conditions into Eq.~\eqref{C4}, it becomes clear
that if there is only a single \textbf{10} Higgs in the model (or, if
$q_{H'}=0$), the requirement of anomaly freedom becomes rather constraining. We
could only find a $\Z8$ symmetry with one generation. However, once we
allow for
non-trivial $q_{H'}$, it is possible to have simple anomaly free \Z{N}
symmetries which satisfy all constraints, the simplest example being a \Z3
symmetry with the charge assignment listed in table~\ref{tab:Z3}. The
$\Z{N\le12}$ symmetries with non-trivial $q_{H'}$ require $N_g=3$ or larger.

\begin{table}
\centerline{\begin{tabular}{|c|c|c|c|c|c|}
\hline
field &  $q_{\psi_m}$ & $H$ & $H'$ & $\Delta$ & $\Delta$\\
\hline
$\Z3$ & 1 & 1 & $-1$ & $-1$ & 1\\
\hline
\end{tabular}}
\caption{\Z3 charge assignment.}
\label{tab:Z3}
\end{table}

Another possible symmetry is $\Z6$ with the charge assignment listed in
table~\ref{tab:SO10}~(c).

This model has the same effective structure as the \textbf{126}
model of
Ref.~\cite{Babu:1992ia,Fukuyama:2002ch,Bajc:2002iw,Goh:2003sy} as
far as the discussion of fermion masses go (even  though it has 2
$\bf{10}$-plets, one  of the $\bf{10}$'s does not couple to
fermions due to $\Z6$ charge assignments).

For the same reasons as in the \textbf{16}-Higgs models, the $Q\,Q\,Q\,L$ type
induced  proton decay operators are forbidden.

Again we refrain from a detailed phenomenological analysis of the way  this
model leads to MSSM at low energies. We close the discussion with a comment on
how easily the MSSM Higgs fields emerge from the  superpotential: the relevant
part of the superpotential has the form:
\begin{eqnarray}
W~=~M_H\, H\,H'+ M_0\,\Delta\, (\overline{\Delta} +\lambda\, \Phi H)+
M_\Phi\,\Phi^2+\lambda'\,\Phi^3 + \lambda''\Delta \Phi H'
+\lambda'''\,\Delta\,\Phi\,\Delta\;,
\end{eqnarray}
which has the right linear combination of MSSM doublets to maintain all the
simple form for the fermion mass results of
Ref.~\cite{Babu:1992ia,Fukuyama:2002ch,Bajc:2002iw,Goh:2003sy}. A detailed
analysis of these issues will be given elsewhere.

\subsection{SO(10) GUTs in higher dimensions}

Let us now comment on implications of our findings for higher-dimensional
models of grand unification, such as `orbifold GUTs'
\cite{Kawamura:1999nj,Kawamura:2000ev,Altarelli:2001qj,Hall:2001pg,Hebecker:2001wq,Asaka:2001eh,Hall:2001xr},
which  provide a simple
solution to the doublet-triplet splitting problem. In such models the dimension
5 proton decay can be naturally suppressed  \cite{Altarelli:2001qj,Hall:2001pg}
(while dimension 6 proton decay is slightly enhanced \cite{Buchmuller:2004eg})
since here the mass partner of the Higgs triplet has vanishing couplings to
matter, as in the discussion above. However, (brane) couplings like $\psi_m^4$,
also leading to proton decay, have not been discussed in this scheme. A reliable
discussion of such operators seems hardly possible in the effective
higher-dimensional field theory framework.

One possible way to address this question is thus to embed the
model into string theory or, in other words, to derive orbifold
GUT models from string theory. The first steps for doing so have
been performed in
Refs.~\cite{Kobayashi:2004ud,Forste:2004ie,Kobayashi:2004ya}. This
has further lead to the scheme of `local grand unification'
\cite{Buchmuller:2004hv,Buchmuller:2005jr,Buchmuller:2006ik,Buchmuller:2007qf},
which facilitates the construction of supersymmetric standard
models from the heterotic string
\cite{Buchmuller:2005jr,Buchmuller:2006ik,Lebedev:2006kn,Lebedev:2007prep}.
Here, the two light MSSM matter generations originate from
$\bf{16}$-plets localized at points with SO(10) gauge symmetry.
Some of these models can have an R-parity arising as \Z2 subgroup
of a gauged, non-anomalous $B-L$ symmetry
\cite{Lebedev:2006kn,Buchmuller:2007zd,Lebedev:2007prep}, like in
ordinary GUTs (however, without the need for
$\overline{\bf{126}}$-plets). On the other  hand, $Q\,Q\,Q\,L$
operators remain a challenge
\cite{Buchmuller:2006ik,Lebedev:2007prep}. The fact that these
operators could be eliminated so easily in conventional GUTs by
simple symmetries leads to the expectation that similar symmetries
will be helpful in the string-derived supersymmetric standard
models with (local) GUT structures. One lesson which one might
learn from our analysis is that one may derive an effective
R-parity and suppress $Q\,Q\,Q\,L$ by a discrete (possibly \Z6)
symmetry under which matter $\bf{16}$-plets have a universal
charge. One might further hope to get insights about the origin of
the discrete symmetries (which remains somewhat obscure in the 4D
field-theoretic approach) in string models. These issues will be
studied elsewhere.

\section{Conclusion and comments}
\label{sec:summary}

Motivated by the beauty of the ideas of supersymmetry and unification, we have
started a search for discrete symmetries that forbid proton decay operators in
gauge extensions of the supersymmetric standard model. We required the
symmetries to allow the standard interactions and to be anomaly free.
Considering the left-right symmetric, Pati-Salam and SO(10) GUT models with
various Higgs contents, we could identify (surprisingly simple) symmetries that
satisfy all our criteria. In many cases, there is a connection between the
anomaly freedom and the number of generations. Often, simple symmetries exist
only for 3 generations (or multiples thereof), as in \cite{Hinchliffe:1992ad}.
In the \SO{10} models, our symmetries forbid dimension 5 proton decay
operators.

Our findings can be interpreted in the following way. Supersymmetric models with
an extended or GUT symmetry are often challenged by proton decay. To rectify
this, one might be forced to introduce additional (discrete) symmetries. Our
examples then show that R-parity can be a consequence of these additional
symmetries rather than being related to $B-L$. From this one might  conclude
that the appearance of fields with even $B-L$ charges is not a necessity, and,
for instance, \textbf{16}-Higgs and \textbf{126}-Higgs SO(10) models can be  on
the same footing.

It is also interesting that the minimal \textbf{126}-based SO(10)  models become
free of all dangerous proton decay operators without  losing their ability to be
predictive in the fermion sector once  we add a simple anomaly free discrete
symmetry.\\[0.1cm]

\noindent\textbf{Acknowledgements.}
We would like to thank Stuart Raby for helpful conversations on proton decay.
This research was supported by the DFG cluster of excellence Origin and
Structure of the Universe and by the SFB-Transregio 27 ``Neutrinos and  Beyond".
The work of RNM is supported by the National Science Foundation grant
no.~Phy-0354401 and by the Alexander von Humboldt Foundation (the Humboldt
Research Award). RNM  would like to thank H.~Fritzsch and A.~Buras for
hospitality at LMU and TUM respectively.

\providecommand{\bysame}{\leavevmode\hbox to3em{\hrulefill}\thinspace}

\end{document}